\documentclass[journal]{IEEEtran}
\usepackage{cite}     
\usepackage{graphicx}

\usepackage{amsmath}

\begin{document}
\title{Triple GEM Detectors for the Forward Tracker in STAR}

\author{{F.~Simon, J.~Kelsey, M.~Kohl, R. Majka, M.~Plesko, D.~Underwood, T.~Sakuma, N.~Smirnov, H.~Spinka and B.~Surrow}
\thanks{Manuskript submitted on \today}
\thanks{F.~Simon is with the Max-Planck-Institut f\"ur Physik, Munich, Germany and with the Excellence Cluster Universe, Technical University Munich, Germany.({\it email: frank.simon@universe-cluster.de}).}
\thanks{ J.~Kelsey, M.~Kohl, M.~Plesko, T.~Sakuma and B.~Surrow are with the Laboratory for Nuclear Science, Massachusetts Institute of Technology. }
\thanks{R.~Majka and N.~Smirnov are with the Physics Department, Yale University.}
\thanks{D.~Underwood and H.~Spinka are with Argonne National Laboratory.}
}

\maketitle

\begin{abstract}
Future measurements of the flavor-separated spin structure of the proton via parity-violating $W$ boson production at RHIC require an upgrade of the forward tracking system of the STAR detector. This upgrade will allow the reconstruction of the charge sign of electrons and positrons produced from decaying $W$ bosons. A design based on six large area triple GEM disks using GEM foils produced by Tech-Etch Inc. has emerged as a cost-effective solution to provide the necessary tracking precision. We report first results from a beam test of three test detectors using Tech-Etch produced GEM foils and a laser etched two dimensional strip readout. The detectors show good operational stability, high efficiency and a spacial resolution of around 70 $\mu$m or better, exceeding the requirements for the forward tracking upgrade. The influence of the angle of incidence of the particles on the spatial resolution of the detectors has also been studied in detail. 
\end{abstract}

\IEEEpeerreviewmaketitle

\section{Introduction}

The study of flavor-separated polarized quark distributions in the proton is one of the cornerstone measurements of the spin physics program with polarized proton collisions at $\sqrt{s}$ = 500 GeV at the Relativistic Heavy Ion Collider. This measurement requires the detection of $W$ bosons through their electron (positron) decay mode. The electrons and positrons can be identified and their energy measured in the electromagnetic calorimeters in STAR. However, the charge sign identification of the outgoing lepton requires high precision tracking which is currently only available at mid rapidity. The identification of the charge of the outgoing lepton at forward rapidity is crucial for this measurement since this provides information on the flavor of the quarks in the initial hard collision. The upgrade of the forward tracker of the STAR experiment \cite{Ackermann:2002ad} at the Relativistic Heavy Ion Collider is a crucial part to achieve the goals of the RHIC Spin program.  In order to identify the charge sign of electrons produced from the decay of $W $ bosons at forward rapidity a multi-layer low mass tracker with \mbox{$\sim$80 $\mu$m} spatial resolution or better is needed. Triple GEM tracking detectors satisfy the requirements for tracking in the forward region in STAR and provide a cost-effective solution.

GEM detectors are based on electron avalanche multiplication in strong electric fields created in holes etched in thin metal clad insulator foils. This concept, introduced in 1996, is referred to as the Gas Electron Multiplier (GEM) \cite{Sauli:1997qp}. Since the electron amplification occurs in the holes in the GEM foil and is separated from charge collection structures, the choice of readout geometries for detectors based on the GEM is very flexible. For tracking applications several GEM foils are cascaded to reach higher gain and high operating stability. Spatial resolutions of better than 70 $\mu$m have been demonstrated with triple GEM detectors \cite{Altunbas:2002ds}, with a material budget of significantly less than 1\% $X_0$ per tracking layer (providing a 2D space point).

\section{The STAR Forward GEM Tracker}

\begin{figure}
\centering
\includegraphics[width=0.48\textwidth]{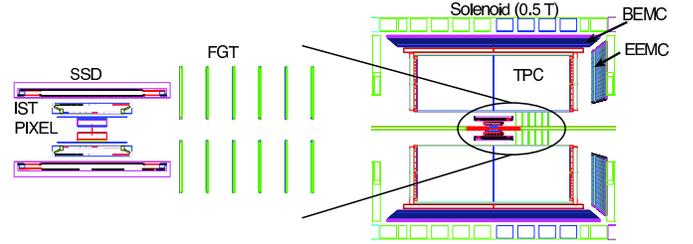}
\caption{Sketch of the STAR tracking upgrade. The Forward GEM Tracker with its 6 triple GEM disks is shown as well as the planned inner silicon tracker, consisting of active silicon pixel sensors and silicon strip detectors.}
\label{fig:FGTView}
\end{figure}

The baseline design of the Forward GEM Tracker FGT consists of 6 triple GEM disks along the beam direction, covering the acceptance of the endcap electromagnetic calorimeter \cite{Allgower:2002zy} for $1< \eta <2$ over the full extend of the interaction diamond in the experiment. The GEM disks will sit inside the inner field cage of the main time projection chamber TPC, and have an outer radius of 38 cm, and an inner radius of 9.5 cm. Figure \ref{fig:FGTView} shows a sketch of the planned forward tracker. In addition, the planned inner tracker upgrade is also shown, as discussed in more detail elsewhere \cite{Simon:2007fz}. The GEM detector disks will be constructed from four quarter sections. This construction requires large area GEM foils. So far the most reliable source of GEM foils is CERN. For a project of that size it is also desirable to have a commercial source of GEM foils in the United States. A collaboration with Tech-Etch, Inc., based on an approved SBIR\footnote{Small Business Innovative Research, US-DOE funded program to foster collaboration of small companies and academic institutions} proposal, has been established to provide a commercial source for GEM foils and to study the production of large area foils. The mechanical construction of the triple GEM disks will be based on light-weight materials such as honeycomb or carbon fiber. A two dimensional projective strip readout will be used. The details of the readout geometry are still under investigation, but are converging towards one strip layer with strips running radially outward, measuring the azimuthal coordinate,with a pitch varying between 300 $\mu$m and 600 $\mu$m, and one layer with azimuthal strips measuring the radial coordinate with a pitch of 800 $\mu$m. The choice of the strip pitch will be discussed later in the test beam data analysis section. The readout system is based on the APV25-S1 front-end chip \cite{French:2001xb}, and will also be used for parts of the silicon tracker upgrade in STAR. Recently the FGT project was reviewed by the Brookhaven National Laboratory Detector Advisory Board and was recommended to be pursued on an aggressive schedule. The total project cost is estimated to be below \$2 million will allow for an accelerated construction of the detector and an installation by fall 2009.

\section{Beam Test Setup}

\begin{figure}
\centering
\includegraphics[width=0.48\textwidth]{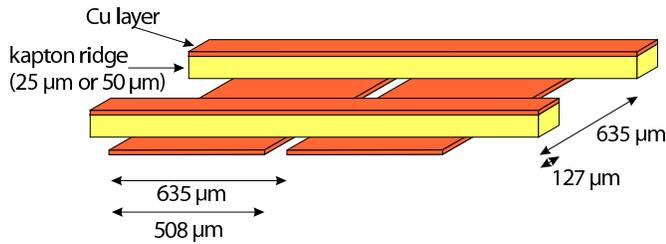}
\caption{The schematic structure of the laser etched 2D orthogonal strip readout board. Two different versions of this board are used, one with 25 $\mu$m and one with 50 $\mu$m thick kapton ridges that define the vertical distance of the two strip layers.}
\label{fig:ReadoutBoard}
\end{figure}

In order to evaluate the performance of GEM foils produced by Tech-Etch in an application environment, a test detector based on the geometry used in the COMPASS experiment \cite{Altunbas:2002ds} has been developed at MIT. The detector is a triple GEM design with a two dimensional projective strip readout. The foils are powered from a single high voltage source through a resistor chain with equal voltage sharing between the three foils. The drift gap of the detector between the cathode foil and the top GEM is 3.2 mm, the transfer gap between the other foils and between the bottom GEM and the readout board are 2.2 mm. The readout structure is a laser etched printed circuit board with a strip pitch of 635 $\mu$m. The top strips are 127 $\mu$m wide and the bottom strips are 508 $\mu$m wide. Two vertical separations of the strip layers are used, one with 25 $\mu$m and one with 50 $\mu$m. This allows the investigation of the effect of the board geometry on the charge sharing between coodinates. The test detectors are designed to allow for easy replacement of individual foils. A pre-mixed gas of Ar:CO$_2$ (70:30) is used for all measurements. Before installation in the detectors all GEM foils are tested for electrical stability and undergo an optical analysis to establish their geometric parametes, using an automated high resolution scanning setup \cite{Becker:2006yc,Simon:2007sk}. Each of the  triple GEM test detectors was also evaluated with a $^{55}$Fe source to study gain uniformity and charging behavior \cite{Simon:2007sk}.

Three triple GEM test detectors with Tech-Etch produced GEM foils were tested in the MTest test beam area at Fermi National Accelerator Laboratory. These detectors where equipped with 6 APV25S1 front-end chips each, with 64 channels per chip connected to readout strips. The APV25 chips where controlled and read out through a custom made module based on FPGAs directly on the detector, which sent its data to a VME unit, which in turn communicated with a DAQ computer via a fiber link developed for the ALICE experiment at CERN. The APV chips were sampling the signal at a rate of 40 MHz and were read out in peak mode, meaning that a single time slice per event was recorded for each strip. The timing of the chips was set up so that this recorded time slice corresponds to the maximum signal amplitude.

The detectors were installed as a tracking telescope with 125 mm spacing between them. The middle detector could be rotated around the vertical axis to study the effect of track inclination on spatial resolution and efficiency. The detectors have a material budget of around 4\% $X_0$ each, mostly due to the bottom support plate made out of aluminum and due to the 2D readout board manufactured on a standard printed circuit board. The effect of multiple scattering on tracks within the GEM telescope was minimized by installing the first detector with its bottom facing the beam, and the other two detectors with their entrance windows facing the beam. That way the amount of material between the active region of the first and the last detector is only around 4\% $X_0$, concentrated in the support plate and the readout board of the central detector. Data was taken under a variety of different beam conditions, with energies ranging from 4 GeV to 32 GeV for unseparated secondary beams, and with a 120 GeV primary proton beam. Over a period of two weeks the detectors were operated without any problems in stable data taking mode. 

\section{First Results}

A set of software tools has been developed to analyze the data taken during the beam test. Preliminary results of this analysis are reported here. 

\begin{figure}
\centering
\includegraphics[width=0.45\textwidth]{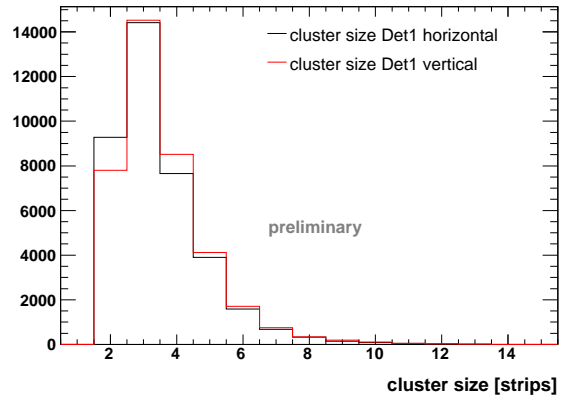}
\caption{The size of reconstructed clusters in strips (635 $\mu$m strip pitch). A minimum width of two strips is imposed by the cluster finder to suppress noise.}
\label{fig:ClusterSize}
\end{figure}

The data is stored in a not zero-suppressed format, with the raw amplitude of each electronics channel available for each event. Since the strip occupancy in the test beam is very low, on the order of 1\% to 3\%, the data events themselves are used to determine the pedestal for each channel on a run-by-run basis. These pedestals are subtracted, and the data is also corrected for common mode noise (correlated shifts of the pedestal for all channels) on a chip-by-chip basis for each event. The corrected channel amplitudes are then used to preform cluster finding, for each of the detector projections separately. The cluster finder is a simple peak finder which seeds the cluster at the strip with the highest amplitude. A minimum of $6\,\sigma_{noise}$ above pedestal, where $\sigma_{noise}$ is the Gaussian sigma of the noise distribution of that particular channel, is required to form a seed. Adjacent strips are added into the cluster as long as their amplitude is more then $1.5\,\sigma_{noise}$ above pedestal. In order to suppress noise from single noisy strips a cluster is required to be at least two strips wide. Figure \ref{fig:ClusterSize} shows the cluster size in the middle detector in the tracking setup. The most likely cluster size is 3 strips on both coordinates. However, it is also apparent that a few clusters are lost due to the requirement of a minimum of two strips. This shows that a smaller strip pitch is desirable to obtain an optimal efficiency and the best possible spatial resolution.

\begin{figure}
\centering
\includegraphics[width=0.495\textwidth]{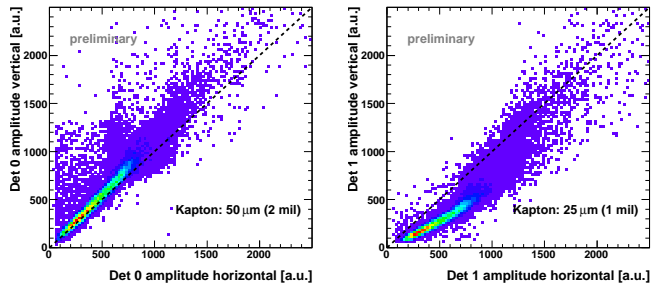}
\caption{Correlation of the reconstructed cluster charge between detector projections. A charge sharing of 1:1 is indicated by the dashed line. The two different readout board designs show different charge sharing between coordinates. While the version with 50 $\mu$m vertical distance between the two strip planes achieves equal charge sharing, the 25 $\mu$m version has a charge sharing of about 1:2 between the top and the bottom strips.}
\label{fig:AmplitudeCorrelations}
\end{figure}

Figure \ref{fig:AmplitudeCorrelations} shows the correlation of the reconstructed cluster charge between both readout coordinates for the two different versions of the readout board that were used in the beam test. The narrow correlation in both cases demonstrate a uniform sharing of the charge over the active area of the detectors. The sharing between coordinates is very different for the two versions of the readout board. While the version with 50 $\mu$m vertical distance between the two strip planes achieves equal charge sharing, the 25 $\mu$m version has a charge sharing of about 1:2 between the top and the bottom strips. This shows that if equal cluster sizes are desired on both coordinates, a design along the lines of the board with 50 $\mu$m vertical distance between strip planes is preferable. 

\begin{figure}
\centering
\includegraphics[width=0.45\textwidth]{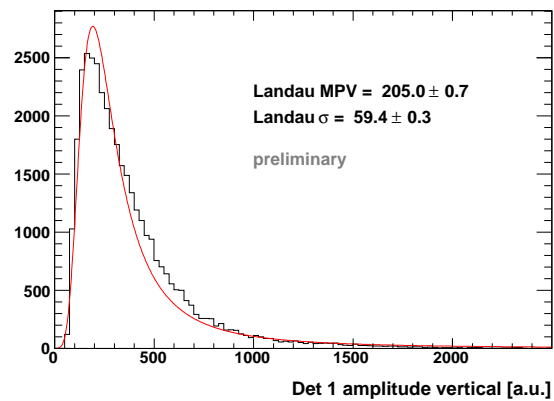}
\caption{Cluster charge distribution on one detector projection, together with a fit with a Landau function. The charge distribution follows the expected Landau shape for thin absorbers.}
\label{fig:AmplitudeLandau}
\end{figure}

Figure \ref{fig:AmplitudeLandau} shows the distribution of the reconstructed cluster charge on one projection of a detector. A Landau function is fitted to the distribution, showing the good agreement with the functional form expected for thin absorbers. The absence of a noise peak at low amplitudes is due to the strict requirements on cluster seed amplitudes and cluster size. 

The spatial position of particles as they traverse the detectors is determined from the weighted mean of the charge recorded on the strips of a cluster. These hits can be used to further study the performance of the detectors. In the preliminary analysis of the data events that have exactly one hit on each projection of the first and the last detector are used to form particle tracks. These are straight lines between the hit positions in the first and last detector. With these tracks the performance of the central detector is investigated in detail. For these studies the data taken with a beam of 32 GeV and of 120 GeV is used to minimize the influence of multiple scattering. The uncertainty of the tracks in the central detector due to multiple scattering is about 11 $\mu$m for 32 GeV and 3 $\mu$m for 120 GeV. 

\begin{figure}
\centering
\includegraphics[width=0.45\textwidth]{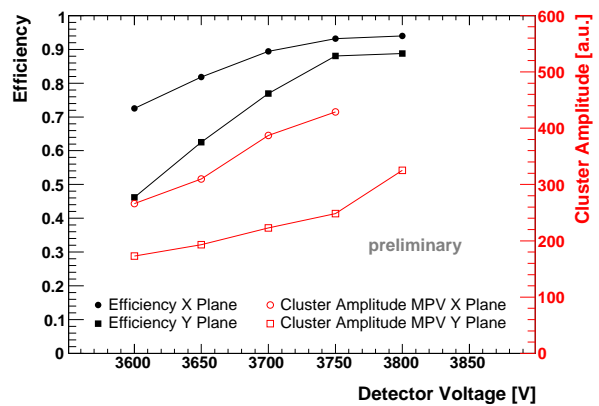}
\caption{Efficiency and cluster amplitude (most probable value obtained by a Landau fit) as a function of detector voltage. For the highest voltage the clusters on the X plane show signs of ADC saturation, thus now reliable amplitude value was obtained.}
\label{fig:EfficiencyAmplitude}
\end{figure}

Figure \ref{fig:EfficiencyAmplitude} shows the single-coordinate efficiency for the middle detector as a function of the applied voltage, obtained for the 32 GeV beam. A hit is counted as identified if it is found within 1 mm of the projected track position, using the same cluster cuts (2 strip minimum) as discussed above. It is apparent that both detectors reach an efficiency plateau at the higher voltages. The Y projection reaches this plateau later due to the 2:1 charge sharing between X and Y. Both projections reach an efficiency of $\sim$90\%, which includes efficiency loss due to dead or noisy strips, and also includes the reduction in efficiency due to the 2 strip minimum requirement for reconstructed clusters, as discussed above and shown in Figure \ref{fig:ClusterSize}. Also shown is the amplitude of the clusters as a function of detector voltage, given by the most probable value of a Landau function fitted to the cluster charge distributions. For the highest voltage the X projection showed signs of ADC saturation, thus no reliable fit was obtained. 

\begin{figure}
\centering
\includegraphics[width=0.45\textwidth]{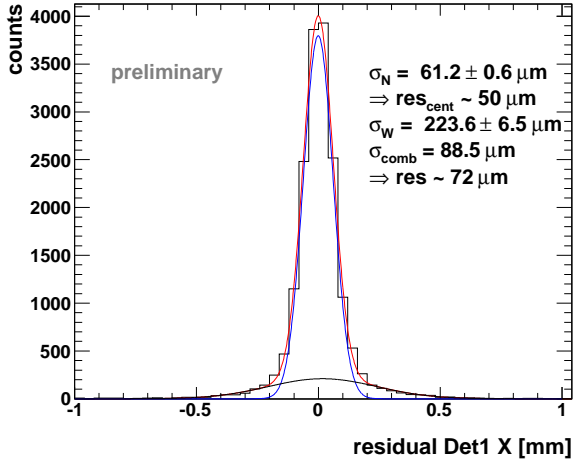}
\caption{Residual distribution of  hits on the X projection of the central detector to tracks formed by the first and last detector in the tracking telescope. The distribution is fitted with the sum of two Gaussian functions, and the spatial resolution of the detectors is extracted assuming equal spatial resolution for all detectors in the telescope.}
\label{fig:ResolutionX}
\end{figure}

From the distribution of the distance of the reconstructed hits in the central detector to the tracks formed by the two outer detectors the spatial resolution of the triple GEM test detectors is extracted. Figure \ref{fig:ResolutionX} shows this residual distribution for the X projection of the central detector for data taken with the 120 GeV beam, where multiple scattering is negligible . The distribution is fitted with the sum of two gaussians, one narrow one for the main central peak of the distribution, and one for the wider shoulders of the distribution. Together this function gives a very good description of the residual distribution. The wider shoulders are probably due to one badly reconstructed hit in the track or in the central detector, for example due to a dead or a noisy strip, or due to very low energy deposit. The width of the central distribution is $\sim$ 61 $\mu$m, while the weighted mean of width of the narrow and wide distribution is around 89 $\mu$m. From the width of the residual distribution the spatial resolution of the detectors is determined by assuming that all three detectors have the same resolution, which is a reasonable assumption since the detectors are identical apart from the different versions of the readout board. In that case the spatial resolution is given by
\begin{equation}
\sigma_{resolution} = \sigma_{residual}\,  \times\, \sqrt{\frac{2}{3}},
\end{equation}
where $\sigma_{residual}$ is the width of the residual distribution determined from the Gaussian fit. With the numbers determined from the fit shown in Figure \ref{fig:ResolutionX} this results in a spatial resolution of $\sim$72 $\mu$m overall and a resolution of $\sim$50 $\mu$m for the main central peak of the distribution. This is well within the requirements for forward tracking in the STAR experiment and demonstrates that resolutions comparable to the ones obtained for the COMPASS triple GEM detectors \cite{Altunbas:2002ds, Ketzer:2004jk} are achievable with detectors based on Tech-Etch produced foils.

In the forward disk configuration under study for the STAR tracking upgrade the angle of incidence for the particle tracks of interest is between $\sim 15^\circ$ and $\sim 30^\circ$. This makes a study of the effect of track inclination on the detector performance crucial. In the test beam this was investigated by rotating the central detector by up to 30$^\circ$ around the vertical axis in the 32 GeV beam. An ionizing particle looses energy in discrete primary interactions with the detector gas. Such interactions can create free electrons, which in turn can ionize other atoms if they are sufficiently energetic. The primary interactions are statistically distributed along the particle track and follow Poissonian statistics in thin gaps. For minimum ionizing particles the density of primary ionizing events is about 4/mm in the Ar:CO$_2$ gas mixture used in the GEM detectors. For a drift gap of 3.2 mm this corresponds to a mean of 13 primary interactions. The energy deposit for each of these primary interactions, and consequently the number of electrons produced at each of these locations, has very large variations. Thus the mean location of all charge produced in the drift gap of the detectors varies widely on an event-by-event basis. The drift and amplification in the detector projects the charge onto the readout board, so for inclined tracks the variation of the mean charge along the particle track directly translates into a variation of the center of gravity of the charge collected on the readout plane, which in turn leads to a deterioration of the spatial resolution. This deterioration affects only the readout projection that is along the inclination of the track, the perpendicular projection is not affected. In the case of the test beam where the detector was rotated around the vertical axis the resolution in the horizontal projection (X) is affected, while the resolution in the vertical projection should remain the same. 

\begin{figure}
\centering
\includegraphics[width=0.45\textwidth]{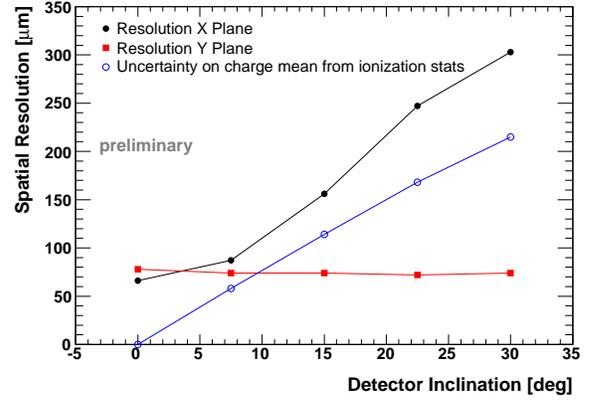}
\caption{Spatial resolution as a function of the angle of incidence of the particle tracks, for both the projection along the inclination of the tracks (X) and the projection perpendicular to the tracks (Y). Also shown is the uncertainty of the center of gravity of the charge due to the ionization statistics obtained from a simulation, as discussed in the text.}
\label{fig:ResolutionAngle}
\end{figure}

Figure \ref{fig:ResolutionAngle} shows the spatial resolution of both projections of the middle detector as a function of the angle of rotation around the vertical axis. As discussed above, the spatial resolution for the X projection deteriorates quickly with increasing angle while the spatial resolution in the Y projection remains unchanged. A simulation based on the techniques described in \cite{Bichsel:2006cs} has been performed to study this effect. The uncertainty of the center of gravity of the charge distribution caused by the statistical distribution of the primary ionization events along the particle track, and the wide spread of energy deposits for each individual interaction also shown in Figure \ref{fig:ResolutionAngle}. This simulation ignores any effects of the strip readout and the digitization, and thus represents a lower limit for the achievable spatial resolution. 

In STAR, the inclination of the particle tracks with respect to the FGT disks will affect the precision of the radial coordinate. The azimuthal coordinate is not affected. In the solenoidal field of the STAR detector the azimuthal coordinate is crucial for the determination of the curvature of the tracks, and thus for the charge sign discrimination necessary for the $W$ program. The fact that the track inclination affects the spatial resolution in one projection has no negative influence on the performance of the STAR forward tracker. The inclination of the particle tracks also leads to a spreading of the charge over a larger area on the readout board. This, in combination with the reduced precision, suggests that a larger strip pitch is sufficient for the radial coordinate. This led to the consideration of a 800 $\mu$m pitch for the strips reading the radial coordinate, while the azimuthal coordinate is determined with strips with a pitch between 300 $\mu$m and 600 $\mu$m. 

\section{Conclusion}

The STAR experiment is preparing an upgrade of its forward tracking system based on triple GEM detectors. Three test detectors using commercially produced GEM foils by Tech-Etch have been tested extensively in a beam test Fermilab. The detectors showed a stable performance during two weeks of beam operations. An efficiency of $\sim$ 90\% including dead and noisy areas and a spatial resolution better than 80 $\mu$m was achieved. Two different versions of a laser-etched 2D orthogonal strip readout board were tested, giving important information on the charge sharing \vadjust{\newpage}between readout coordinates depending on the board geometry. Overall the results of the beam test demonstrate that devices using commercially produced GEM foils from Tech-Etch satisfy the requirements of forward tracking in the STAR experiment at RHIC. Currently the design of the large-area detectors is being finalized and the proposal for the forward tracking upgrade in STAR is being reviewed.

\section*{Acknowledgments}

The authors thank Fermilab for the allocation of beam time and Eric Ramberg and colleagues for generous support during the test beam activities. The development of GEM foil production at Tech-Etch is supported by US-DOE SBIR grant DE-FG02-05ER84169.

\bibliographystyle{IEEEtran.bst}
\bibliography{GEM}

\begin{thebibliography}{10}
\providecommand{\url}[1]{#1}
\csname url@rmstyle\endcsname
\providecommand{\newblock}{\relax}
\providecommand{\bibinfo}[2]{#2}
\providecommand\BIBentrySTDinterwordspacing{\spaceskip=0pt\relax}
\providecommand\BIBentryALTinterwordstretchfactor{4}
\providecommand\BIBentryALTinterwordspacing{\spaceskip=\fontdimen2\font plus
\BIBentryALTinterwordstretchfactor\fontdimen3\font minus
  \fontdimen4\font\relax}
\providecommand\BIBforeignlanguage[2]{{%
\expandafter\ifx\csname l@#1\endcsname\relax
\typeout{** WARNING: IEEEtran.bst: No hyphenation pattern has been}%
\typeout{** loaded for the language `#1'. Using the pattern for}%
\typeout{** the default language instead.}%
\else
\language=\csname l@#1\endcsname
\fi
#2}}

\bibitem{Ackermann:2002ad}
K.~H. Ackermann \emph{et~al.}, ``{STAR} detector overview,'' \emph{Nucl.
  Instrum. Meth.}, vol. A499, pp. 624--632, 2003.

\bibitem{Sauli:1997qp}
F.~Sauli, ``{GEM}: A new concept for electron amplification in gas detectors,''
  \emph{Nucl. Instrum. Meth.}, vol. A386, pp. 531--534, 1997.

\bibitem{Altunbas:2002ds}
M.~C. Altunbas \emph{et~al.}, ``Construction, test and commissioning of the
  triple-{GEM} tracking detector for {COMPASS},'' \emph{Nucl. Instrum. Meth.},
  vol. A490, pp. 177--203, 2002.

\bibitem{Allgower:2002zy}
C.~E. Allgower \emph{et~al.}, ``The {STAR} endcap electromagnetic
  calorimeter,'' \emph{Nucl. Instrum. Meth.}, vol. A499, pp. 740--750, 2003.

\bibitem{Simon:2007fz}
F.~Simon, ``The {STAR} tracking upgrade,'' \emph{arXiv:0710.0172
  [physics.ins-det]}, 2007.

\bibitem{French:2001xb}
M.~J. French \emph{et~al.}, ``Design and results from the {APV25}, a deep
  sub-micron {CMOS} front-end chip for the {CMS} tracker,'' \emph{Nucl.
  Instrum. Meth.}, vol. A466, pp. 359--365, 2001.

\bibitem{Becker:2006yc}
U.~Becker, B.~Tamm, and S.~Hertel, ``Test and evaluation of new {GEMs} with an
  automatic scanner,'' \emph{Nucl. Instrum. Meth.}, vol. A556, pp. 527--534,
  2006.

\bibitem{Simon:2007sk}
F.~Simon \emph{et~al.}, ``Development of tracking detectors with industrially
  produced {GEM} foils,'' \emph{arXiv:0707.2543 [physics.ins-det]}, 2007.

\bibitem{Ketzer:2004jk}
B.~Ketzer, Q.~Weitzel, S.~Paul, F.~Sauli, and L.~Ropelewski, ``Performance of
  triple {GEM} tracking detectors in the {COMPASS} experiment,'' \emph{Nucl.
  Instrum. Meth.}, vol. A535, pp. 314--318, 2004.

\bibitem{Bichsel:2006cs}
H.~Bichsel, ``A method to improve tracking and particle identification in
  {TPC}s and silicon detectors,'' \emph{Nucl. Instrum. Meth.}, vol. A562, pp.
  154--197, 2006.

\end{thebibliography}

\end{document}